# Computing Power, Key Length and Cryptanalysis. An Unending Battle?

Aristides Dasso[a], Ana Funes[a], Daniel Riesco[a], Germán Montejano[a]

[a]*SEG, Universidad Nacional de San Luis,*
*Ejército de Los Andes 950, D5700HHW*
*San Luis, Argentina*
*{arisdas, afunes,driesco,gmonte}@unsl.edu.ar*

2018

## Abstract

*There are several methods to measure computing power. On the other hand, Bit Length (BL) can be considered a metric to measure the strength of an asymmetric encryption method. We review here ways to determine the security, –given an span of time– of a factoring-based encryption method –such as RSA– by establishing a relation between the processing power needed to break a given encryption and the given bit length used in the encryption. This relation would help us provide an estimation of the time span that an encryption method for a given BL will be secure from attacks.*

## 1. Introduction

For our purposes we define a document as a text, image, sound, etc., that needs to be transmitted or kept. In both cases it has to be secured against inspection, modification, or both, by non authorised agents; an agent being a person or a computer program. The document's protection has a life span, i.e. a period of time during which the protection should work.

Along history there has been a need for protection against potential attacks to a document. This need has been satisfied by storing the document in a secure place or using encryption of the document. Several encryption methods have been used, from the most simple transposition of symbols or even the use of a different language, to the more modern methods used in electronic transmission and storage, so is safely to say that encryption has a long history that precede their use in the computer area.

Nowadays we can distinguish two main families of encryption methods, namely Symmetric and Asymmetric (or Public-Key). Symmetric methods are used from what can be called the beginnings of cryptography to the present. Symmetric methods use the same cryptographic key for both encryption and decryption of a document.

Asymmetric (or Public-Key) methods rely –at least at present– on what they are called 'one way functions'. This concept can be exemplified by quoting W. S. Jevons, who said in 1874: "Can the reader say what two numbers multiplied together will produce the number 8,616,460,799 ? I think it unlikely that anyone but myself will ever know ; for they are two large prime numbers, and can only be rediscovered by trying in succession a long series of prime divisors until the right one be fallen upon. The work would probably occupy a good computer for many weeks, but it did not occupy me many minutes to multiply the two factors together." [[12] ]. However, according to [[21] ] the Jevons' number was factored by Bancroft Brown around 1925.

Without going into the theory of 'one way functions', it can safely being said that a one-way function is a function that is 'easy' to compute for any input, but 'hard' to compute the inverse, i.e. given an output of the function is hard to obtain the corresponding input. 'Easy' and 'hard' relate to computational complexity theory.

Going back to Jevons' example, it is not 'impossible' to find the two factors that produce 8,616,460,799, is 'hard', and getting easier –less 'hard' if you will– with the increase along time of electronic computer power. One of the most commonly used methods is based on what is known as the factorization problem, i.e. how long it takes to factor a composite number.

A document that is encrypted has a life span; this life span is related to the nature of the document and this is also related to the security of the encryption method. For instance, nowadays, a document that is a message which only needs to be secure until electronic delivery, has a very short life span, probably measured in seconds or fractions of a second; while a document that is digitally signed using an encryption method would need a life span of several years. The protection during the life span of a document is related –when the encryption method is based on the factorization problem– on the time it takes to find the factors of a composite number (public key) and this is related to the length (number of digits) of the composite number to be factored. Hence determining the required life span of a document is directly related to the number of digits of the encryption key. At present, the length of an encryption key is measured in bits (BL).

So time comes now into the picture, i.e. life span of a document as well as time taken to break a given encryption key. So, to keep a secret based on the 'hardness' of a computation at a given moment and pretending to be so kept for a period of time, then the ongoing increase in computer power as well as cryptanalysis must be considered. That is the gist of this article, namely, if at time $t_0$ a document is normally encoded with $x_0$ bits and at time $t_1$ encoding is normally done using $x_1$ bits ($x_0 < x_1$), can a document encoded with $x_0$ bits be broken at time $t_1$?, and if so, how large will $x_0$ have to be; at least $x_0 = x_1$?, or larger still?, and how to estimate, when you are at $t_0$, the value of $x_0$?.

The rest of the article is organized as follows: in section 2 a short presentation of the RSA method is given. RSA is the Asymmetric (or Public-Key) method that we have analyzed, since RSA is currently not only the most used method but also the method on which there are more attacks done. In section 3 the different attacks that RSA has suffered are reviewed and compared to some statistics on the increase in computer power along the years. In section 4 we present a method to combine time, BL and computer effort to estimate the BL needed at a particular time period. In Section 5 we discuss the data presented and its possible outcomes. Section 6 gives some conclusions.

**Table 1. Bit lengths of various cryptosystems for different security levels [[22] ]**

| Algorithm Family | Crypto systems | Security Level (bit) | | | |
|---|---|---|---|---|---|
| | | 80 | 128 | 192 | 256 |
| Integer factorization | RSA | 1024 bit | 3072 bit | 7680 bit | 15360 bit |
| Discrete logarithm | DH, DSA, Elgamal | 1024 bit | 3072 bit | 7680 bit | 15360 bit |
| Elliptic curves | ECDH, ECDSA | 160 bit | 256 bit | 384 bit | 512 bit |
| Symmetric-key | AES, 3DES | 80 bit | 128 bit | 192 bit | 256 bit |

## 2. Asymmetric Encryption Methods - RSA

The example given above by Jevons [[12] ] is actually the basis of one of the most used method nowadays of public key cryptography, i.e. RSA. Discovery of the basics of the public key method is normally attributed to Whitfield Diffie and Martin Hellman [[8] ] although others –Ralph Merkle [[18] ], James Ellis, Clifford Cocks, etc., see [[29] ] for more on this– can claim to have discovered a similar method.

The RSA method was actually developed by Rivest, Shamir and Adleman [[24] ] in 1978. RSA stands for the initials of their last names.

RSA encryption works as follows: given a public key $k_{pub} = (n, e)$, and a plain document represented by a number $M$, the encryption function that produced the encrypted document $C$, is:

$$C = e(M) \equiv M^e \bmod n$$
where:
$$n = p \bullet q; (p, q) \text{ are primes}$$
$$\text{and } M, C \in \mathbb{Z}_n$$
(1)

RSA decryption works as follows: given the private key $k_{pr} = d$ and the cipher text $C$,

$$M = d(C) \equiv C^d \bmod n$$
where:
$d$ is a large integer, that is relatively prime to
$(p - 1) \bullet (q - 1)$,
which is the same as
$\Phi(n) = (p - 1) \bullet (q - 1)$
i.e. $\gcd(d, (p - 1) \bullet (q - 1)) = 1$
and $M, C \in \mathbb{Z}_n$
(2)

While the encryption key $(n, e)$ is public, i.e. known to everybody; the decryption key, $d$, is private; also $(p, q)$, are known only to the proprietor of the decryption key.

$n$ is sometimes known as the modulus, and since $n$ is the product of two prime numbers ($p$ and $q$ in equation (1)); therefore, the security of RSA is based on the difficulty of the integer factorization problem.

## 3. Attacks on RSA

The security level of a cryptosystem can be measured in bits. In general, "An algorithm is said to have a "security level of $b$ bit if the best known attack requires $2^b$ steps." [[22] ]. Following this, an example of an estimation of security level is given in Table 1.

The size of an RSA key refers either to the bit-length of the RSA modulus or the number of decimal digits of the number ($n$ in equation (1) and **¡Error! No se**

**encuentra el origen de la referencia.**)). So in RSA–<*number*>, <*number*> refers to the number of decimal digits, or in other cases to the number of bits. Chiefly, decimal digits were used from RSA-100 to RSA-500, and bits from RSA-576 on [[31] ].

RSA Laboratories had a challenge to encourage research into the difficulty of factoring large integers. The challenge ran from 28 January 1997 until May 2007. A list of the numbers proposed for the challenge with the author, date, etc., on which some of them were factored, can be found now on [[32] ].

### 3.1. RSA Attacks that are not contemplated.

There are different attacks on RSA security to obtain the private key, ranging from social engineering to straight stealing of the key, bad management of the key and their generation, among others. However we are not concerned here with these kinds of attacks on RSA. We considered these attacks as 'physical' attacks and their security is correspondingly 'physical' and do not depend

**Table 2. Computing Time Effort. [[10] ] [[14] ]**

| Number | Date | MIPS /year | Algorithm | Extrapolation from: RSA-140 | RSA-130 |
|---|---|---|---|---|---|
| C116 | 1990 | 275 | MPQS | | |
| RSA120 | 1993 | 830 | MPQS | | |
| RSA129 | 1994 | 5,000 | MPQS | | |
| RSA130 | 1996 | 1,000 | NFS | | |
| RSA140 | 1999 | 2,100 | NFS | | |
| RSA155 | 1999 | 8,400 | NFS | 16,800 | 33,600 |

on the Bit Length of the key.

We do not considered either the feasibility of obtaining the key using methods like eavesdropping with sophisticated means such as those described in [[11] ].

Also, the possibility of different organizations having factored large integers –larger than are publicly known at the time of writing– and keeping it private. However, on this last point, [[17] ] believes that "unpublished work is many years ahead of what the public at large gets to see". But, since, as was said above, we are not considering this scenario, neither will we estimate how many years ahead is the unpublished work.

Also the advent of quantum computing and corresponding algorithms such as [[27] ] are not contemplated; although progress continues in this area [[7] ].

Excluded as well are different proposals of special hardware to perform factorization [[25] ] [[26] ] [[28] ].

Whether factoring a large integer is easy or harder than breaking RSA –or solving the so called RSA problem– is also not covered here, for more on this see [[1] ] [[2] ] [[3] ] [[4] ].

### 3.2. RSA Attacks analyzed.

The attacks on RSA that we considered are those based on the factorization of large integers, i.e. getting p and q from n; see equation (1) and **¡Error! No se encuentra el origen de la referencia.**).

If an attacker succeeds in factoring n, then it can get d –and then M, of course– as shown in equation (3).

$$\begin{array}{ll} 1^{st} & \Phi(n) = (p-1) \bullet (q-1) \\ 2^{nd} & d^{-1} \equiv e \bmod \Phi(n) \qquad (3) \\ 3^{rd} & M \equiv C^d \bmod n \end{array}$$

Up to the time of writing, the claim for the largest known RSA number factored is RSA-768 [14] . In [[32] ] 

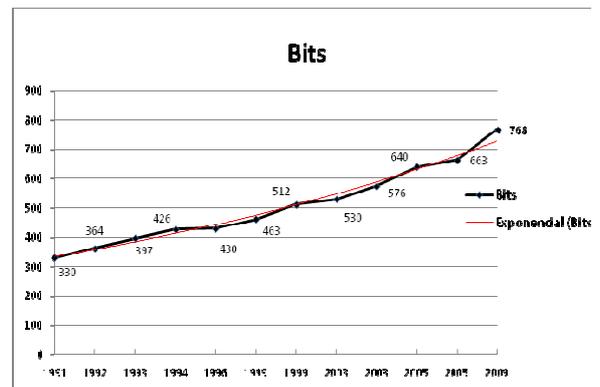

**Figure 1. RSA bit length / date factored**

the RSA Numbers, the date when they were factored and who did it, can be found.

Figure 1 shows a curve where the RSA key length is plotted against the date when it was factored. It also shows the tendency as an exponential line.

Not only is it important the key length in the factoring process but also the cost of said process. Cost can be measured in several ways, from simple monetary cost, ranging to speed, time, etc., obviously the complexity of the algorithm used, as stated above, is the absolute cost – at least for the time being– in terms of the computing effort. However, from the point of view of security, overall speed in breaking a given RSA number is one important factor, if not the most important. Monetary cost on the other hand could be said to be directly related to the expected monetary gain in breaking a given RSA key.

Along history there have been many methods to try integer factorization, from the sieve of Eratosthenes, to Fermat's factorization algorithm. It can be said that John

M. Pollard new method, i.e. Pollard's ρ algorithm (1975), is much faster than trial division. MPQS, invented by Carl Pomerance in 1981 was a significant step, and then of course NFS in 1993.

Currently factorization is done using the Number Field Sieve (NFS) [[15]], or General Number Field Sieve; that was an important improvement from previous methods, e.g.: Multiple Polynomial Quadratic Sieve (MPQS). NFS considerably accelerated the factoring process.

**Table 3. RSA number, year factored and hours taken for factoring**

| Number | Year | Hours taken for factoring | |
|---|---|---|---|
| RSA-512 | 1999 | 5,040 [[5]] | |
| RSA-512 | 2015 | 4.[30] | 1,260 times less than RSA-512 (1999) |
| RSA-768 | 2009 | 21,600 [[14]] | 4.29 times more than RSA-512 (1999) |

This change can be observed in Table 2 where there was a factor of 5 in improvement in effort measured in MIPS/year when changing from MPQS to NFS. Also, the extrapolations from factoring the previous RSA numbers –RSA-130 and RSA-140– to the effort that factoring RSA-155 should have taken, shows differences of 33,6 and 2 times respectively with the actual effort taken. This reveals that the effort needed –in this case measured in MIPS/year– not only decreases along time, something to be expected, but it does so dramatically.

The complexity of factoring an RSA number can be expressed, using *L*-notation, by equation (4); that is the heuristic complexity associated with NFS [[15]]. It is not a polynomial time, and is not either an exponential one; it's considered super-polynomial. The algorithm in [[27]] is considered polynomial.

$$L[n, u, c] = exp(c(\ln n)^u (\ln \ln n)^{1-u})$$
Were $c = (64/9)^{1/3} \approx 1.923$ and $u = 1/3$ (4)
For convenience $L[n, u, c]$ is referred as $L[n]$

The fastest factoring of an RSA number is claimed by [[30]]; they factored RSA-160, 512 bits, in 2015, in under four hours using Amazon Elastic Compute Cloud platform for a cost of $75; and although they were not the first by a long shot, the great reduction in processing time is a significant advance and a clear indication of the importance of cloud computing and must be taking into consideration for the future of the security of factoring based systems.

## 4. Key length estimation

Key length (BL), given in bits, is critical for the security of a cryptographic protocol. But BL has to be considered together with the time span that the security is desired to be kept.

There are several points to be considered when estimating the strength in time of a given key, namely, processing power, theoretical computational effort needed –see equation (4) above– that implies not only the amelioration of the current method but also the appearance of a radically new method, hardware cost, empirical data from the published reports of RSA breaking. The theoretical computational effort needed – see equation (4) above– is unlikely to change in the foreseeable future unless a new method is introduced.

On the other hand, is difficult to predict the advent of a radically new method [[17]]. The arrival of a totally new factoring method that transform factoring from superpolynomial to polynomial –e.g. a quantum method– unpublished reports coming to light, or anything equally catastrophic from the security point of view, etc., as was already said above, cannot be considered since it is nearly impossible to estimate anything about them and it would amount to sheer speculation. Should any of this happen, then all bets are off.

So in the next sections we will be considering only three factors, namely experimental data on factoring RSA numbers, computer power, and factoring effort based on equation (4).

### 4.1. Data on factoring and their relation to each other

It is also true that the actual method employed, namely GNFS, is steadily being improved, but these improvements do not produce significant changes as a completely new method will.

Table 3 shows three RSA number with the year they were factored, and the number of hours taken to do it. We can see that the two RSA-512 were factored 16 years apart, but the second time (2015) took approximately 1,260 times less hours than the first (1999). RSA-768, that is 1.5 times longer than RSA-512, took approximately 4.3 times more (in 2009) than RSA-512 (1999).

### 4.2. Moore´s law and BL estimation

The increase in computer power is closely related by many –amongst others [[9] ] [[13] ] [[16] ] [[17] ]– to the so called Moore´s law [[19] ] [[20] ] that although applies to the density of components in an integrated circuit, it has been correlated with computer power in general. Initially, according to Moore this density doubles every 18 months, so computer power should double every 18 months. Whether this doubling will keep in the future is a matter for debate. Several times it has been predicted its dismissal, however it continues to be 'alive', and there are indications that it will continue to be so, since new fabrication methods continue appearing [[6] ].

**Table 4. Doubling every 18m vs. real time taken**

| Number | Year factored | Months in between | Doubling (months) | Hours |
|---|---|---|---|---|
| RSA-512 | 1999 | | | |
| RSA-512 | 2015 | 192 | 18 | 3.10059 |
| RSA-512 | 2015 | 192 | 18.6422 | 4 |

**Table 6. Time estimation for factoring RSA-768, RSA-1024 and RSA-2048**

| Number | Time taken (hours) | Time Estimation | |
|---|---|---|---|
| | | hours | years |
| RSA-512 | 4 | | |
| RSA-768 | 21,600 | 24,567 | 2.80 |
| RSA-1024 | | 29,989,314 | 3,423 |
| RSA-2048 | | 3.50954805E+16 | 4.00633E+12 |

This assumption –increase in computer power and Moore's law– fits in closely enough with the increase in the factoring speed of a RSA number, namely RSA-512. If 1999 is chosen as a starting period, there are 192 months between 1999 –the first time RSA-512 was factored– and 2015 –the second time RSA-512 was factored–; if computational power would have double every 18 months (there are 10.67 periods of 18 months in the 192 months, with a total doubling of $2^{10.67}$ times), then factoring in 2015 should have taken 3.10059 hours (5,040 hs / $2^{10.67}$) –all things being equal– instead of the actual 4 hours reported, so this means doubling every 18.6422 months (192 / $\log_2(5,040 / 4)$) instead of every 18 months; this is shown in Table 4. In 1975 Moore [[20] ] gives an estimation of 2 years instead of the previous 1.5, but we believe that the 1.5 years could be kept for our purposes given the previous analysis based on the numbers in Table 4.

The differences in these numbers are small enough to attribute them to small errors and therefore they can be ignored, and accept –conservatively– the 18 months doubling for the computational power increase. Since the factoring of RSA-512 in 2015 was done using cloud computing we can say that the doubling takes care of the use of cloud computing too.

**Table 5. Effort for RSA-512, RSA-768, RSA-1024 and RSA-2048, using equation (4)**

| Number | Effort (*L*) | Times harder |
|---|---|---|
| RSA-512 | 1.73671977E+19 | |
| RSA-768 | 1.06663813E+23 | 6.14168242E+03 |
| RSA-1024 | 1.30207587E+26 | 7.49732856E+06 |
| RSA-2048 | 1.52377537E+35 | 8.77387012E+15 |

Considering the experimental data coming from the actual time taken to factorize RSA numbers, we can see that this data succeeds in explaining both the increase in computer power as well as the steady enhancement of the current factoring method, i.e. GNFS.

### 4.3. Complexity measure and estimation

We used equation (4) to calculate the effort required for factoring RSA-512, RSA-768, RSA-1024 and RSA-2048. This is shown in Table 5, where how much harder factoring RSA-768, RSA-1024 and RSA-2048 than RSA-512, is also presented.

Table 6 shows the claimed time taken to factor RSA-512 in 2015 [[30] ] and RSA-768 in 2009 [[14] ] and also, for RSA-768, RSA-1024 and RSA-2048, an estimation of the time it would take given how much harder each one is in relation to RSA-512 (2015).

The numbers in Table 5 could be an overestimation of the effort needed. However this estimation is not too far out considering that the numbers shown in Table 6 for RSA-768 are not too different (only approximately 14% more) than the actual time taken. When considering estimation, for a cryptographic key for security reasons, is not too bad to err on the cautious side. This overestimation can increase as *n* grows bigger. Again, since we are trying to come out with a BL that gives a reasonable guaranteed of protection against attacks, we are not overly concern with it.

**Table 7. Breaking RSA-512 considering doubling power 1.5 years**

| Year | years | minutes |
|---|---|---|
| 2015 | | 240 |
| 2016,5 | 1.5 | 120 |
| 2018 | 3 | 60 |
| 2021 | 6 | 30 |
| 2027 | 12 | 15 |
| 2039 | 24 | 7,5 |
| 2063 | 48 | 3,75 |
| 2111 | 96 | 1,875 |

## 5. Discussion

Looking at Table 7 we can see that considering the data of the breaking of RSA-512 in 2015 and supposing a doubling of computer power every 1.5 years, by the year 2039, just 22 years away (from 2018), it would take seven and a half minutes to break an 512-encryption.

On the other hand, when taking into account the effort needed –as presented in Table 5–, if we look at Table 8, that shows the time it would take factoring RSA-1024 and RSA-2048 if we suppose that RSA-512 would take 1 minute (instead of the four hours it really took to break it), then RSA-1024 would take more than 14 years to be broken and RSA-2048 would take more than 16 billion (1.66930558E+10) years to be broken. All this at present conditions, of course.

In [[23] ] it is recommended that the minimum size for an encryption up to the year 2030 should be 2048, while 3072 bits (or more) should be used after the year 2030. "RègleFact-1. La taille minimale du module est de 2048 bits, pour une utilisation ne devant pas dépasser l'année 2030. RègleFact-2. Pour une utilisation au-delà de 2030, la taille minimale du module est de 3072 bits.".

However, for anything that goes beyond a horizon fifty years away is difficult to ascertain its reliability. Who can say that a new polynomial algorithm or machine –whether quantum or otherwise– cannot appear in fifty years time? And that raises an important issue, namely, an encoding, e.g. for a digital signature, that should last one, or even two generations –say that we consider a 'generation' to be twenty five years– and also if we take into consideration that whoever signed the document could not be around anymore to change the signature of the document, then there is no guaranteed that a 2048 or even larger BL would last that long to protect a digital signature.

## 6. Conclusion

We have considered BL for factoring based encryption methods to evaluate the BL that could give the best possible protection during a particular time frame, especially when the time desired time span extend for several years. To achieve that we have resorted to analyze decryption methods, focusing our attention on factoring algorithms that are the main breaking methods used at present.

Table 8. Considering RSA-512 would take 1 minute

| Number | Time Estimation | | |
|---|---|---|---|
| | minutes | hours | years |
| RSA768 | 6,141.68 | 102.3613737 | 0.01 |
| RSA-1024 | 7,497,328.56 | 124,955.4761 | 14.26 |
| RSA-2048 | 8.77387012E+15 | 1.46231E+14 | 1.66930558E+10 |

As was said above, we have not considered social engineering methods since they belong more to 'physical' security protection mechanisms. Neither were analyzed the possibilities of new and better factoring algorithms that those currently in use, or the appearance of different type of computers such as quantum or otherwise. It is very hard to estimate the speed of the advances in those two areas.

In analysing the breaking of factoring based encryption methods we directed our attention to RSA that is one of the most widely used method and for which there is plenty of evidence of successful breaking attempts. Also we employed the heuristic formula for the effort needed to factorize different key lengths. So using both data –'real life' breaking of different RSA lengths, as reported in the literature and the effort needed, for a given RSA length, at the time of breaking– we give an estimation of the BL needed to keep an encryption secured for an specified time frame.

So a question is: will a new polynomial algorithm or machine appears in the next fifty years (two generations in our previous scenario)? If the answer is yes then there is no BL that would protect a document encoded with a factoring–based algorithm. If the answer is not then the other alternative is to choose a BL of 2048, that considering the present situation, and equation (4), should be good enough to ensure security for the next 16 billion years.